\documentclass{aastex631}
\usepackage[T1]{fontenc}

\usepackage{graphicx}
\usepackage{txfonts}
\usepackage{natbib}
\bibpunct{(}{)}{;}{a}{}{,} 
\usepackage{xcolor}
\usepackage{multirow}

\usepackage{svg}

\newcommand{\orcid}[1]{\href{https://orcid.org/#1}{\includesvg[width=10pt]{orcid}}}
\DeclareSymbolFont{UPM}{U}{eur}{m}{n}
\SetSymbolFont{UPM}{bold}{U}{eur}{b}{n}
\DeclareMathSymbol{\upi}{0}{UPM}{"19}
\DeclareMathSymbol{\umu}{0}{UPM}{"16}

\newcommand{\orcit}[1]{\protect\href{https://orcid.org/#1}{\protect\includegraphics[width=8pt]{orcid.png}}}

\usepackage{orcidlink}
\begin{document} 

   \title{A close binary lens revealed by the microlensing event \textbf{Gaia20bof}}
     \author[0000-0002-6578-5078]{E. Bachelet} 
    \affiliation{IPAC, Mail Code 100-22, Caltech, 1200 E. California Blvd., Pasadena, CA 91125, USA;}
    
    \author{P.~Rota}
    \affiliation{ Dipartimento di Fisica ``E.R. Caianiello'', Universit\`a di Salerno, Via Giovanni Paolo II 132, 84084 Fisciano, Italy}
    \affiliation{  Istituto Nazionale di Fisica Nucleare, Sezione di Napoli, Via Cintia, 80126 Napoli, Italy}
   
    \author{V.~Bozza}
    \affiliation{ Dipartimento di Fisica ``E.R. Caianiello'', Universit\`a di Salerno, Via Giovanni Paolo II 132, 84084 Fisciano, Italy}
    \affiliation{  Istituto Nazionale di Fisica Nucleare, Sezione di Napoli, Via Cintia, 80126 Napoli, Italy}  
    
    \author[0000-0001-6434-9429]{P. Zieli{\'n}ski}
    \affiliation{ Institute of Astronomy, Faculty of Physics, Astronomy and Informatics, Nicolaus Copernicus University in Toru{\'n}, Grudzi\k{a}dzka 5, 87-100 Toru{\'n}, Poland}

    \author[0000-0001-8411-351X]{Y. Tsapras}
    \affiliation{Zentrum f{\"u}r Astronomie der Universit{\"a}t Heidelberg, Astronomisches Rechen-Institut, M{\"o}nchhofstr. 12-14, 69120 Heidelberg, Germany}

    \author[0000-0003-0961-5231]{M.~Hundertmark}
    \affiliation{Zentrum f{\"u}r Astronomie der Universit{\"a}t Heidelberg, Astronomisches Rechen-Institut, M{\"o}nchhofstr. 12-14, 69120 Heidelberg, Germany}

    \author{J.~Wambsganss}
    \affiliation{Zentrum f{\"u}r Astronomie der Universit{\"a}t Heidelberg, Astronomisches Rechen-Institut, M{\"o}nchhofstr. 12-14, 69120 Heidelberg, Germany}

    \author[0000-0002-9658-6151]{{\L}. Wyrzykowski}
    \affiliation{Astronomical Observatory, University of Warsaw, Al.~Ujazdowskie~4, 00-478~Warszawa, Poland}

    \author{P. J. Miko{\l}ajczyk}
    \affiliation{Astronomical Observatory, University of Warsaw, Al.~Ujazdowskie~4, 00-478~Warszawa, Poland}
    \affiliation{Astronomical Institute, University of Wroc{\l}aw, Kopernika 11, 51-622 Wroc{\l}aw, Poland}
    
    \author{R.A.~Street} 
    \affiliation{Las Cumbres Observatory Global Telescope Network, Inc., 6740 Cortona Drive, Suite 102, Goleta, CA 93117, USA}

    \author{R. Figuera Jaimes}
    \affiliation{ Millennium Institute of Astrophysics MAS, Nuncio Monsenor Sotero Sanz 100, Of. 104, Providencia, Santiago, Chile}
    \affiliation{Instituto de Astrof\'isica, Facultad de F\'isica, Pontificia Universidad Cat\'olica de Chile, Av. Vicu\~na Mackenna 4860, 7820436 Macul, Santiago, Chile}

    \author{A.~Cassan}
    \affiliation{Institut d'Astrophysique de Paris, Sorbonne Universit\'e, CNRS, UMR 7095, 98 bis bd Arago, F-75014 Paris, France}

    \author{M.~Dominik}
    \affiliation{University of St Andrews, Centre for Exoplanet Science, School of Physics \& Astronomy, North Haugh, St Andrews, KY16 9SS, United Kingdom}
    
    \author{ D.~A.~H.~Buckley}
    \affiliation{South African Astronomical Observatory, PO Box 9, Observatory Rd, Observatory 7935, South Africa}
    \affiliation{Department of Astronomy, University of Cape Town, Private Bag X3, Rondebosch 7701, South Africa}
    \affiliation{Department of Physics, University of the Free State, PO Box 339, Bloemfontein 9300, South Africa}

    \author{S.~Awiphan}
    \affiliation{National Astronomical Research Institute of Thailand (Public Organization), 260 Moo 4, Donkaew, Mae Rim, Chiang Mai 50180, Thailand}

    \author{N. Nakhaharutai}
    \affiliation{Department of Statistics, Faculty of Science, Chiang Mai University, Chiang Mai 50200, Thailand}

    \author{S. Zola}
    \affiliation{Astronomical Observatory, Jagiellonian University, Orla 171, 30-244 Krak{\'o}w, Poland}
    
    \author{ K. A. Rybicki}
    \affiliation{Astronomical Observatory, University of Warsaw, Al.~Ujazdowskie~4, 00-478~Warszawa, Poland}
    \affiliation{Department of Particle Physics and Astrophysics, Weizmann Institute of Science, Rehovot 76100, Israel}

    \author{M. Gromadzki}
    \affiliation{Astronomical Observatory, University of Warsaw, Al.~Ujazdowskie~4, 00-478~Warszawa, Poland}

    \author{K. Howil}
    \affiliation{Astronomical Observatory, University of Warsaw, Al.~Ujazdowskie~4, 00-478~Warszawa, Poland}

    \author{ N. Ihanec}
    \affiliation{Astronomical Observatory, University of Warsaw, Al.~Ujazdowskie~4, 00-478~Warszawa, Poland}
    
    \author{ M. Jab{\l}o{\'n}ska}
    \affiliation{Astronomical Observatory, University of Warsaw, Al.~Ujazdowskie~4, 00-478~Warszawa, Poland}
    \affiliation{Research School of Astronomy \& Astrophysics, Australian National University, Cotter Rd., Weston, ACT 2611, Australia}

    \author{ K. Kruszy{\'n}ska}
    \affiliation{Astronomical Observatory, University of Warsaw, Al.~Ujazdowskie~4, 00-478~Warszawa, Poland}
    \affiliation{Las Cumbres Observatory Global Telescope Network, Inc., 6740 Cortona Drive, Suite 102, Goleta, CA 93117, USA}

    \author{ K. Kruszy{\'n}ska}
    \affiliation{Astronomical Observatory, University of Warsaw, Al.~Ujazdowskie~4, 00-478~Warszawa, Poland}
    \affiliation{Las Cumbres Observatory Global Telescope Network, Inc., 6740 Cortona Drive, Suite 102, Goleta, CA 93117, USA}

    \author{U. Pylypenko}
    \affiliation{Astronomical Observatory, University of Warsaw, Al.~Ujazdowskie~4, 00-478~Warszawa, Poland}

    \author{M. Ratajczak}
    \affiliation{Astronomical Observatory, University of Warsaw, Al.~Ujazdowskie~4, 00-478~Warszawa, Poland}

    \author{M. Sitek}
    \affiliation{Astronomical Observatory, University of Warsaw, Al.~Ujazdowskie~4, 00-478~Warszawa, Poland}

    \author{M. Rabus}
    \affiliation{Departamento de Matem{\'a}tica y F{\'i}sica Aplicadas, Facultad de Ingenier{\'i}a, Universidad Cat{\'o}lica de la Sant{\'i}sima Concepci{\'o}n, Alonso de Rivera 2850, Concepci{\'o}n, Chile}

    \date{Received ??; accepted ??}
 
\begin{abstract}
   {During the last 25 years, hundreds of binary stars and planets have been discovered towards the Galactic Bulge by microlensing surveys. Thanks to a new generation of large-sky surveys, it is now possible to regularly detect microlensing events across the entire sky.}
   {The OMEGA Key Projet at the Las Cumbres Observatory carries out automated follow-up observations of microlensing events alerted by these surveys with the aim of identifying and characterizing exoplanets as well as stellar remnants. In this study, we present the analysis of the binary lens event Gaia20bof. }
   {By automatically requesting additional observations, the OMEGA Key Project obtained dense time coverage of an anomaly near the peak of the event, allowing characterization of the lensing system.}
   {The observed anomaly in the lightcurve is due to a binary lens. However, several models can explain the observations. Spectroscopic observations indicate that the source is located at $\le2.0$ kpc, in agreement with the parallax measurements from Gaia.}
   {While the models are currently degenerate, future observations, especially the Gaia astrometric time series as well as high-resolution imaging, will provide extra constraints to distinguish between them.}
\end{abstract}

   
%
\section{Introduction}
The gravitational microlensing effect \citep{Einstein1936} has been used for more than twenty years to detect faint objects in the Milky Way. Originally used to probe the nature of the dark matter in the galactic halo \citep{Paczynski1986} by observing toward the Magellanic Clouds, microlensing surveys are now focused on the Galactic Bulge, where the event rate is the highest. Thanks to the Optical Gravitational
Lensing Experiment \citep{Udalski2015}, the Microlensing Observation in Astrophysics \citep{Sumi2003}, the Korea Microlensing Telescope Network \citep{Kim2016} and follow-up teams\footnote{a summary can be found \url{http://www.microlensing-source.org/follow-up-programs/}}, more than hundred exoplanets have been detected\footnote{according to the NASA Exoplanet \url{https://exoplanetarchive.ipac.caltech.edu/}}. More recently, the first isolated stellar-mass black hole has been detected by microlensing using precise astrometry from the Hubble Space Telescope \citep{Sahu2022,Lam2022, Mroz2022}.

A new generation of large-sky surveys opens the possibility to detect microlensing across the entire sky. The combination of large field-of-view, high spatial resolution, low limiting magnitudes and rapid data processing offer the possibility to discover microlensing event in the Galactic Disk. This opens the opportunity to study the galactic population of lenses that are difficult to observe otherwise. In particular, the Gaia mission of the European Space agency has detected hundreds of microlensing events \citep{Wyrzykowski2023}. Events observed by Gaia are often of special interest for several reasons. The duration of events in the Galactic Disk is usually longer, allowing the detection of additional effects such as the microlensing parallax \citep{Gould1994} or the orbital motion of the lens \citep{Wyrzykowski2020}, which provide unique constraints on the mass and distances of the lenses. Moreover, the future astrometric time series delivered by the Gaia mission offers the possibility to measure the angular Einstein ring radius $\theta_\mathrm{E}$ via the astrometric microlensing signal \citep{Rybicki2018}, providing a mass/distance relation for a large fraction of lenses. By design, these surveys generally deliver a weekly cadence of observation, which is not dense enough to accurately catch anomalous features in microlensing events. Therefore, the follow-up of these events is of paramount importance to ensure their characterisation \citep{2018Geosc...8..365T}.

 As with most microlensing events detected by the Gaia mission, Gaia20bof (equatorial: ($\alpha=184.61816^{\circ}$, $\delta=-63.49726^{\circ}$, J2000), galactic ($l=299.26406^{\circ}$,$b=-0.86052$)) is located in the Galactic Disk. The event was announced by the Gaia Science Alerts (GSA \footnote{\url{http://gsaweb.ast.cam.ac.uk/alerts/alert/Gaia20bof/}} hereafter) \citep{Hodgkin2021} on 30 March 2020 and the microlensing nature of the event was confirmed via spectroscopic classification from Southern African Large Telescope, SALT \citep{Ihanec2020ATel}, on the 10 June 2020, because the data displays a stellar spectrum with only absoprtion lines. In response to this alert, several groups started to observe this event and identified an anomalous feature at the event peak. We present the different observations collected for this event in Section~\ref{sec:data}. The modelling of the photometric observations presented in Section~\ref{sec:modelling} reveals that the lightcurves can be equally well explained by different competing models. Section~\ref{sec:source} presents the analysis of the different spectra collected and confirms the measurement from Gaia revealing that the source is a red subgiant located at ~2\,kpc. As discussed in Section~\ref{sec:future} and Section~\ref{sec:conclusion}, this implies that the lens is relatively close and that additional follow-up observations will place additional constraints on the models in the near future.
   
\section{Observations}\label{sec:data}

\subsection{Gaia}
Gaia's photometric measurements consist of a wide $G$-band \citep{Jordi2010} obtained with roughly monthly cadence while Gaia scans the sky. The photometry is publicly available on the GSA web page \citep{Hodgkin2021}. We used the procedure described in \citet{Kruszynska2022} to obtain the photometric errors of the 163 measurements used in this study. 

\subsection{The OMEGA Key Project}
The OMEGA Key Project is an international collaboration that performs automatic follow-up of microlensing events detected by large-sky surveys. The primary goal is the characterization of cold planets and stellar remnants in the entire sky. Microlensing candidates are first collected from various channels, including from GSA, in the Microlensing Observing Platform (MOP)\footnote{\url{https://mop.lco.global/}}. The MOP system is a Target and Observation Manager, or TOM system, built with the TOM Toolkit package \citep{Street2018}. As soon as data are available, this system automatically fits a single lens model (including microlensing parallax, see for example \citet{Gould2004}) to all of the data available for each event, using the pyLIMA modelling software \citep{Bachelet2017}. Typically, there are more potential targets at any given time than there are telescope resources to observe them and it is therefore necessary to prioritize targets. MOP incorporates an algorithm that prioritize targets based on their position in the sky and their current status \citep{Hundertmark2018}. For instance, events located towards the galactic Bulge ($255^{\circ}\le\alpha\le275^{\circ}$ and $-36^{\circ}\le\delta\le-22^{\circ}$) are observed only if they are sufficiently sensitive to planets according to \citet{Hundertmark2018}. This is motivated by the fact that this region is regularly monitored by OGLE, MOA and KMTNet surveys. Otherwise, higher cadence imaging and spectroscopic observations are requested automatically by the MOP system, tailored to current parameters and phase of the evolution of each event($\sim$ daily). Most of the observations are collected via the Las Cumbres Observatory (LCO) automatic robotic telescopes network \citep{Brown2013}. The observing strategy consists of regular monitoring of the event in two bands, namely SDSS-g' and SDSS-i' (with the exception of events located towards the Galactic Bulge, where the SDSS-g' band is replaced by the SDSS-r' band, due to the higher extinction in these fields), with a cadence depending of the event priority and Einstein ring crossing time $t_E$. Observing in two bands serves two primary purposes. First, it measures the chromaticity of ongoing events. Indeed, microlensing events are achromatic phenomena (to first order) and therefore the color evolution contributes to exclude astrophysical false-positive detections (such as Be stars) but also helps to distinguish between microlensing models (namely, the double source/double lens scenario, see below). Second, as emphasized by \citet{Yoo2004}, observations in (at least) two bands allows the  estimation of the angular source radius, a key component in estimating the mass of the lenses. We note that the cadence in the SDSS-i' band is twice that of the other bands in order to increase the sampling of the lightcurve. If at any given time, models predict a high planet sensitivity (during a high magnification event for example) \citep{Hundertmark2018}, a 15-minute cadence mode is triggered for the next 48 hours to ensure that any potential anomalies are well sampled. If a target is predicted to exceed a brightness threshold of 17 mag, several low-resolution spectra (R$\sim$1000) using the FLOYDS instruments \citep{Brown2013} are also requested, to help with the source characterization, see for example \citet{Fukui2019} as well as to classify contaminants in the alerts stream, such Young-Stellar Objects. If the event gets bright enough (i.e. $V\le11$ mag), high-resolution NRES spectra (R$\sim$55000) \citep{Siverd2018} can also be triggered to obtain a precise estimation of the source spectral parameters, such as in the event Gaia19bld \citep{Bachelet2022}.

As listed in Table~\ref{tab:datasets}, hundreds of images have been collected in the SDSS-g' and SDSS-i' bands, from La Silla in Chile (LSC), Siding Spring in Australia (COJ) and the South African Astronomical Observatory in South Africa (CPT) LCO sites. Some of these images come from precursor observing programs prior to OMEGA. We note that, with the exception of the LSC site that remained closed for a long time, the COVID-19 pandemic only mildly impacted this observing program at LCO.  

\subsection{Other follow-up data}
The target has been followed up photometrically with a network of small telescopes under the umbrella of the OPTICON-RadioNET Pilot program of the European Commission, e.g.  \cite{Wyrzykowski2020}. For Gaia20bof we used LCO, TRT and Skynet's PROMPT-MO and PROMPT5 telescopes, the summary of these observations is gathered in Table \ref{tab:datasets}.

TRT stands for Thai Robotic Telescopes, a 0.7m telescope equipped with  Andor iKon-L DW936 BV CCD camera with resolution 0.8 arcseconds/pixel. The telescope is located at 259.3045 deg West and 
26.6955 deg North.

PROMPT-MO is a 0.4 m RC telescope equipped with Apogee USB CCD with resolution 0.598 arcseconds/pixel.
The telescope is located at 243.011 deg West and 31.638 deg South in the Meckering Observatory, Australia.

PROMPT5 is another Skynet Network telescope and is a Ritchey-Chretien 0.41 m telescope operating using Apogee CCD camera at 0.8 arcseconds/pixel resolution. 
The telescope is located in CTIO Chile at 70.8053889 deg West and 30.1676389 deg South. 

\begin{table*}

\setlength\extrarowheight{3pt}
\caption{Summary of the observations.}
    \label{tab:datasets}
    \centering
    \begin{tabular}{lcc}
    \hline
    \hline
    
         Name & Filter & Observations  \\
         \hline
         
         Gaia & G & 163 \\
         LCO\_V & V & 6 \\
         LCO\_I & I & 6 \\
         LCO\_gp & SDSS-g' & 36 (Rybicki/ORP)+161(Bachelet/Omega) \\
         LCO\_ip & SDSS-i' & 40 (Rybicki/ORP)+221(Bachelet/Omega) \\
         TRT\_I & I & 45 \\
         TRT\_V & V & 44 \\
         PROMPT\_MO\_B$^\mathrm{a}$ & B & 1 \\
         PROMPT\_MO\_I$^\mathrm{a}$ & I & 8 \\
         PROMPT\_5\_B$^\mathrm{a}$ & B & 2 \\
         PROMPT\_5\_I$^\mathrm{a}$ & I & 7 \\
         PROMPT\_5\_gp$^\mathrm{a}$ & SDSS-g' & 6 \\
         \hline
         Total & & 746 \\
    \hline     
    \end{tabular}
    
a. These observations were taken solely at the baseline and hence were not used for the modelling.
\end{table*}




\begin{figure*}
    \centering
    \includegraphics[width=0.9\textwidth]{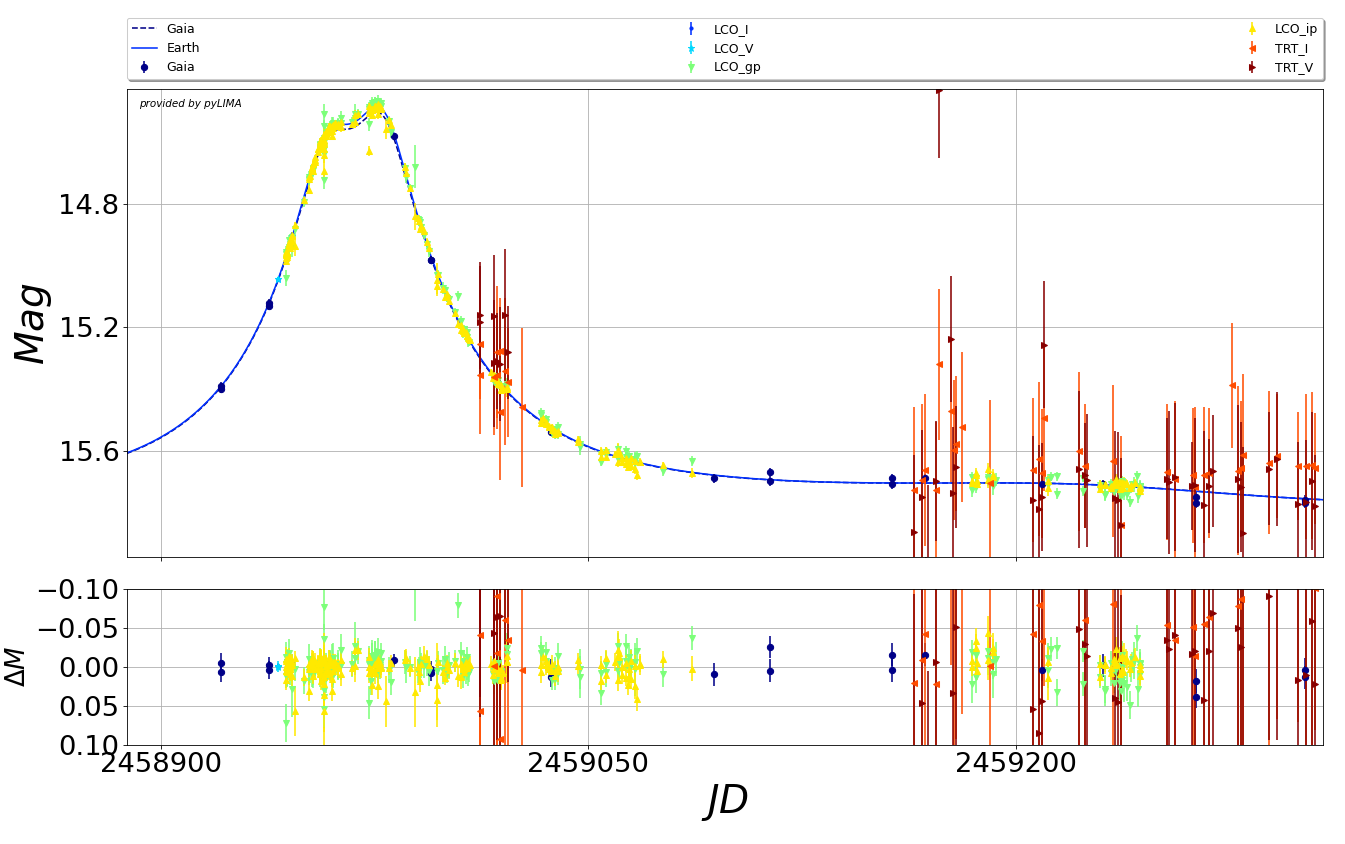}
    \caption{ Lightcurves of the event observed from the ground and Gaia. The Wide+ model is presented (blue solid line, the dashed line represents the model seen by Gaia) but all models produce almost identical lightcurves. The parallax signal between Gaia and Earth is almost not visible. All datasets have been artificially shifted to Gaia magnitudes for presentation purposes. The presented uncertainties take into account the rescaling parameters derived during the posterior exploration, see the main text for more details.}
    \label{fig:lightcurve}
\end{figure*}

\subsection{Spectroscopic data}
\label{sec:spectro}
We collected two spectra during the course of the event: one is from 10-m Southern African Large Telescope (SALT; \citealt{Buckley2006}) equiped with the Robert Stobie Spectrograph \citep{Burgh2003,Kobulnicky2003}, and the second is from the X-Shooter intrument \citep{Vernet2011} mounted on the ESO 8-m Very Large Telescope (VLT). 

The SALT/RSS low-resolution spectrum was obtained on 2020 June 6 (JD$\sim$2459006) at event magnification $\sim$1.8\footnote{SALT Large Programme ID: 2018-2-LSP-001, PI: D. Buckley}, with the exposure time 600 s. The longslit mode was used with the slit width 1.5 arcsec, grating pg0300. The wavelength range of the obtained spectrum covers 370-930~nm giving the resolving power $R\sim350$ and average signal-to-noise ratio S/N=160. It has been reduced in a standard way using PySALT\footnote{\url{http://pysalt.salt.ac.za/}} software (bias subtraction and flat-field correction; \citealt{Crawford2010}) and then wavelength and flux calibration was applied using standard IRAF routines thanks to having calibrating data for Ar comparison lamp and spectrophotometric standard star.

The VLT/X-Shooter spectrum was obtained at the baseline of the microlensing event on January 7, 2022\footnote{ESO Programme ID: 108.22JZ.001, PI: {\L}. Wyrzykowski} for three wavelength channels: UVB ($300-559.5$~nm), VIS ($559.5-1024$~nm) and NIR ($1024-2480$~nm). We integrated with 191 s, 220 s and 300 s of exposure time for UVB, VIS and NIR channel, respectively. The resolution of this spectrum is $R\sim10000$ in UVB part (at slit width 1.0 arcsec) and $R\sim15000$ in VIS (at slit width 0.7 arcsec) and NIR (at slit width 0.6 arcsec) parts, on average. It has been reduced with the dedicated EsoReflex\footnote{\url{https://www.eso.org/sci/software/esoreflex/}} pipeline (v. 2.9.1). For the calibration of UVB, VIS wavelengths, ThAr lamp was used, while for NIR -- a set of Ar, Hg, Ne and Xe lamps. Due to the poor quality of some spectral ranges and low signal-to-noise ratio, only the parts between $350-552$~nm, $560-745$~nm, $780-914$~nm and $1133-1350$~nm, $1450-1800$~nm, $1950-2356$~nm were used in further analysis.

\section{Data reduction and modelling}
\label{sec:reduction}

\subsection{Photometric reductions}
\label{sec:photo}
The follow-up observations were reduced using the Black Hole TOM (BHTOM) infrastructure\footnote{\url{http://bhtom.space}}, which utilised CCDPhot suite of image processing and CPCS photometry calibration tools\citep{Zielinski2019, Zielinski2020}. The photometry was calibrated to the Gaia Synthetic Photometry catalog Gaia synthetic SDSS magnitudes \citep{Montegriffo2022}. We note that the magnitudes are in the AB system \citep{Oke1983}.

\subsection{Modelling of microlensing event}
\label{sec:modelling}
As can be seen in Figure~\ref{fig:lightcurve}, the OMEGA follow-up data clearly reveal an anomaly around $\mathrm{JD} \sim 2458970$. First, we explored the binary lens versus the binary source interpretation \citep{Dominik2019} and found that the latter can be safely ruled out because of the high $\chi^2$ value (about two times higher than the binary lens solutions). However, the anomaly is mild, with no clear sign of a caustic crossing and looks similar to the approach of a Chang-Refsdal lens \citep{Chang1979}. Such lightcurves are notoriously known for presenting strong degeneracies in terms of lens geometries, see for instance \citet{Dominik1999,Han2008,Shvartzvald2016,Bozza2016}, implying that several models can reproduce the observations accurately, which is exactly the case here. A binary model is parametrised by the time $t_0$ of the minimum impact parameters $u_0$ relative to the centre of mass of the binary lens. The angular Einstein ring radius crossing time is defined as $t_\mathrm{E}=\frac{\theta_\mathrm{E}}{\mu_\mathrm{{rel}}}$, where $\mu_\mathrm{rel}$ is the (geocentric) relative proper motion. We also considered $\rho= \theta_\star/\theta_\mathrm{E}$, the normalised source radius where $\theta_*$ is the angular source radius, in the models, but this parameter is not constrained in the present case, because the source trajectories are far from the caustic. For this reason, we did not consider limb-darkening effects. The models also include the normalised angular projected separation between the two components of the lens $s$, and their mass ratio $q$. Finally, the angle between the lens trajectory and the binary axes is defined by $\alpha$ (counter-clockwise). 
Because of the event duration ($t_\mathrm{E} \ge 50~\mathrm{d}$), we also consider the microlensing parallax vector $\mathbf{\pi_E} = (\pi_{\mathrm{EN}},\pi_{\mathrm{EE}})$ \citep{Gould2004} and set the time of reference $t_{0,\mathrm{par}}=t_0$ for all models. The norm of the parallax vector is directly related to the lens and source distances via:
\begin{equation}
||\mathbf{\pi_E}||={{\pi_{LS}}\over{\theta_E}}    
\end{equation}
with $\pi_{LS} = 1/D_l-1/D_s$, while the parallax vector is co-linear to the lens-source relative proper motion projected in the North-East plan of the sky.
The first steps of modelling have been done using the \mbox{RTModel} infrastructure\footnote{\url{https://www.fisica.unisa.it/gravitationastrophysics/RTModel.htm}}, which located eight minima. We refine these solutions with a Monte-Carlo Markov Chain (MCMC) exploration on each of them as reported in the Table~\ref{tab:Models}. The posterior exploration has been performed with the updated version 2.0 of pyLIMA \citep{Bachelet2017}, which maximizes the total log-likelihood $\cal L$:
\begin{equation}
{\cal L} = -0.5 \sum_{n=1}^{N_\mathrm{telescopes}} \sum_{t=1}^{N_\mathrm{data}} \left [{{(f_{n,t}-m_{n,t})^2}\over{\sigma_{n,t}^2}}+\ln\,(2\pi \sigma_{n,t}^2)\right]
\end{equation}
where $N_\mathrm{telescopes}$ is the total number of telescopes, $N_\mathrm{data}$ is the total number of photometric observations of a given telescope, $\sigma_t$ are the uncertainties associated with the measured flux $f_t$ at the time $t$ and $m_t$ is the corresponding microlensing model flux.
The version 2.0 of pyLIMA introduces the option of rescaling the flux uncertainties $\sigma_{i,t}$ of each telescope $i$ during the MCMC exploration using
\begin{equation}
\sigma_{i,t}' = 10^{k_i}\sigma_{i,t}\,.
\end{equation}
A detailed presentation of the new features available with the latest release of pyLIMA will be the subject of a separate paper (Bachelet et al., in prep.). The MCMC exploration is performed with the \texttt{emcee} package \citep{ForemanMackey2013} with a fixed number of 10000 chains and 32 walkers. The convergence of the chains have been analyzed separately for each model. A summary of the best fitting models found is displayed in Table~\ref{tab:Models}, and the best model is presented in Figure~\ref{fig:lightcurve}. For each parameter set, the 16, 50 and 84 percentiles of the chains are shown. We note that the reported $\chi^2$ values for each model are obtained from a gradient fit started at the best MCMC chain-position (and includes the rescaled uncertainties obtained from the chains). The Wide+ model (i.e. $s\ge1$) is currently slightly favoured and points towards a stellar binary lens interpretation, but the planetary companion scenario remains possible at this time.

\begin{longrotatetable}
\label{tab:Models}
\begin{deluxetable*}{cccccccccc}
\tablecaption{Best models from the modelling of the lightcurves. The microlensing parameters and errors corresponds to the 16, 50 and 84 percentiles of the MCMC chains. The $\chi^2$ corresponds to the maximum-likelihood and are based on rescaled uncertainties rescaling.}
\tabletypesize{\small}
\tablehead{
\colhead{Parameters} 
&\colhead{Unit}&\colhead{CloseB-}&\colhead{CloseB+}&\colhead{CloseP-}&\colhead{CloseP+}&\colhead{Resonant-}&\colhead{Resonant+}&\colhead{Wide-}&\colhead{Wide+}}
\startdata
$t_0$&${\rm JD}$&$2458971.83_{-0.17}^{+0.16}$&$2458971.90_{-0.17}^{+0.23}$&$2458969.458_{-0.075}^{+0.074}$&$2458969.360_{-0.068}^{+0.068}$&$2458969.427_{-0.053}^{+0.066}$&$2458969.342_{-0.065}^{+0.069}$&$2458827.7_{-59.9}^{+63.5}$&$2458847.2_{-4.2}^{+9.7}$ \\
$u_0$&&$-0.2744_{-0.0093}^{+0.0085}$&$0.2552_{-0.0073}^{+0.0065}$&$-0.29_{-0.03}^{+0.03}$&$0.225_{-0.010}^{+0.012}$&$-0.306_{-0.015}^{+0.014}$&$0.254_{-0.012}^{+0.011}$&$-0.12_{-0.02}^{+0.02}$&$1.66_{-0.10}^{+0.11}$ \\
$t_\mathrm{E}$&${\rm days}$&$56.6_{-1.5}^{+1.2}$&$63.9_{-1.2}^{+1.6}$&$55.0_{-2.6}^{+3.0}$&$67.9_{-2.2}^{+1.8}$&$53.9_{-1.9}^{+1.9}$&$69.9_{-2.5}^{+3.2}$&$492.7_{-98.9}^{+174.5}$&$133.8_{-3.8}^{+3.7}$ \\
$\rho$&&$0.038_{-0.026}^{+0.031}$&$0.017_{-0.014}^{+0.024}$&$0.15_{-0.08}^{+0.02}$&$0.055_{-0.024}^{+0.019}$&$0.1808_{-0.0068}^{+0.0067}$&$0.064_{-0.030}^{+0.019}$&$0.015_{-0.014}^{+0.015}$&$0.037_{-0.020}^{+0.014}$ \\
$s$&&$0.4178_{-0.0060}^{+0.0075}$&$0.4327_{-0.0054}^{+0.0065}$&$0.661_{-0.103}^{+0.047}$&$0.586_{-0.013}^{+0.017}$&$1.013_{-0.023}^{+0.029}$&$1.424_{-0.037}^{+0.033}$&$3.604_{-0.044}^{+0.043}$&$4.082_{-0.076}^{+0.075}$ \\
$q$&&$1.05_{-0.09}^{+0.11}$&$0.84_{-0.08}^{+0.07}$&$0.025_{-0.007}^{+0.018}$&$0.0374_{-0.0031}^{+0.0029}$&$0.0181_{-0.0017}^{+0.0025}$&$0.0475_{-0.0032}^{+0.0035}$&$0.770_{-0.043}^{+0.053}$&$1.40_{-0.12}^{+0.12}$ \\
$\alpha$&${\rm rad}$&$0.911_{-0.020}^{+0.018}$&$5.368_{-0.024}^{+0.019}$&$1.643_{-0.005}^{+0.006}$&$4.6527_{-0.0037}^{+0.0039}$&$1.6452_{-0.0050}^{+0.0044}$&$4.6517_{-0.0037}^{+0.0046}$&$0.209_{-0.099}^{+0.071}$&$5.155_{-0.012}^{+0.021}$ \\
$\pi_{\rm EN}$&&$0.368_{-0.016}^{+0.015}$&$0.227_{-0.010}^{+0.013}$&$0.333_{-0.024}^{+0.022}$&$0.25_{-0.01}^{+0.01}$&$0.341_{-0.013}^{+0.013}$&$0.243_{-0.010}^{+0.011}$&$0.059_{-0.040}^{+0.025}$&$0.082_{-0.008}^{+0.013}$ \\
$\pi_{\rm EE}$&&$-0.346_{-0.013}^{+0.016}$&$-0.353_{-0.012}^{+0.014}$&$-0.40_{-0.02}^{+0.02}$&$-0.320_{-0.013}^{+0.014}$&$-0.408_{-0.020}^{+0.021}$&$-0.311_{-0.014}^{+0.015}$&$-0.2894_{-0.0095}^{+0.0088}$&$-0.306_{-0.011}^{+0.011}$ \\\\
\hline
$\cal L$&&$-6048$&$-6050$&$-6052$&$-6056$&$-6048$&$-6057$&$-6062$&$-6044$ \\
$\chi^2/{\rm dof}$&&720/692&726/692&707/692&721/692&709/692&707/692&723/692&703/692 \\\\
\hline
$G_\mathrm{S}$&&$15.905_{-0.050}^{+0.043}$&$16.005_{-0.037}^{+0.043}$&$15.87_{-0.11}^{+0.11}$&$16.152_{-0.071}^{+0.057}$&$15.821_{-0.075}^{+0.074}$&$16.156_{-0.073}^{+0.091}$&$16.095_{-0.057}^{+0.076}$&$16.052_{-0.043}^{+0.044}$ \\
$G_\mathrm{B}$&&$18.18_{-0.29}^{+0.51}$&$17.60_{-0.17}^{+0.17}$&$18.1_{-0.5}^{+1.3}$&$17.1_{-0.1}^{+0.2}$&$18.8_{-0.7}^{+1.1}$&$17.1_{-0.2}^{+0.2}$&$17.29_{-0.20}^{+0.19}$&$17.44_{-0.14}^{+0.17}$ \\
$g_\mathrm{S}$&&$17.055_{-0.048}^{+0.037}$&$17.129_{-0.034}^{+0.039}$&$17.02_{-0.11}^{+0.11}$&$17.265_{-0.067}^{+0.056}$&$16.960_{-0.071}^{+0.072}$&$17.267_{-0.068}^{+0.086}$&$17.237_{-0.050}^{+0.066}$&$17.209_{-0.042}^{+0.043}$ \\
$g_\mathrm{B}$&&$19.22_{-0.23}^{+0.41}$&$18.80_{-0.15}^{+0.18}$&$19.3_{-0.5}^{+1.2}$&$18.4_{-0.1}^{+0.2}$&$20.0_{-0.7}^{+1.1}$&$18.4_{-0.2}^{+0.2}$&$18.36_{-0.14}^{+0.14}$&$18.51_{-0.15}^{+0.15}$ \\
$i_\mathrm{S}$&&$15.435_{-0.048}^{+0.037}$&$15.510_{-0.034}^{+0.039}$&$15.40_{-0.11}^{+0.11}$&$15.645_{-0.067}^{+0.056}$&$15.340_{-0.071}^{+0.072}$&$15.648_{-0.068}^{+0.086}$&$15.616_{-0.051}^{+0.066}$&$15.589_{-0.042}^{+0.043}$ \\
$i_\mathrm{B}$&&$17.25_{-0.17}^{+0.28}$&$16.93_{-0.12}^{+0.14}$&$17.41_{-0.46}^{+0.98}$&$16.56_{-0.12}^{+0.16}$&$17.89_{-0.53}^{+0.91}$&$16.56_{-0.17}^{+0.17}$&$16.58_{-0.13}^{+0.12}$&$16.70_{-0.12}^{+0.12}$ \\ \\
\hline
$\theta_*$ & $\mathrm{\umu as}$ & $5.5\pm0.2$&$5.3\pm0.2$&$5.1\pm0.2$&$5.4\pm0.2$&$5.1\pm0.2$&$5.0\pm0.2$&$5.5\pm0.2$&$5.0\pm0.2$ \\ \\ 
\enddata
\end{deluxetable*}
\end{longrotatetable}

\subsection{Spectroscopic properties of the source}
\label{sec:source}
Based on the low-resolution SALT/RSS spectrum we were able to classified the source as a reddened GK-type star \citep{Ihanec2020ATel}. Data shows the stellar spectrum with only absorption lines, therefore, the changes of the brightness of Gaia20bof could be explained by the gravitational microlensing phenomenon. Therefore, we decided to continue with extensive photometric follow-up monitoring of Gaia20bof event. 

\begin{figure*}
    \centering
    \includegraphics[width=0.95\textwidth]{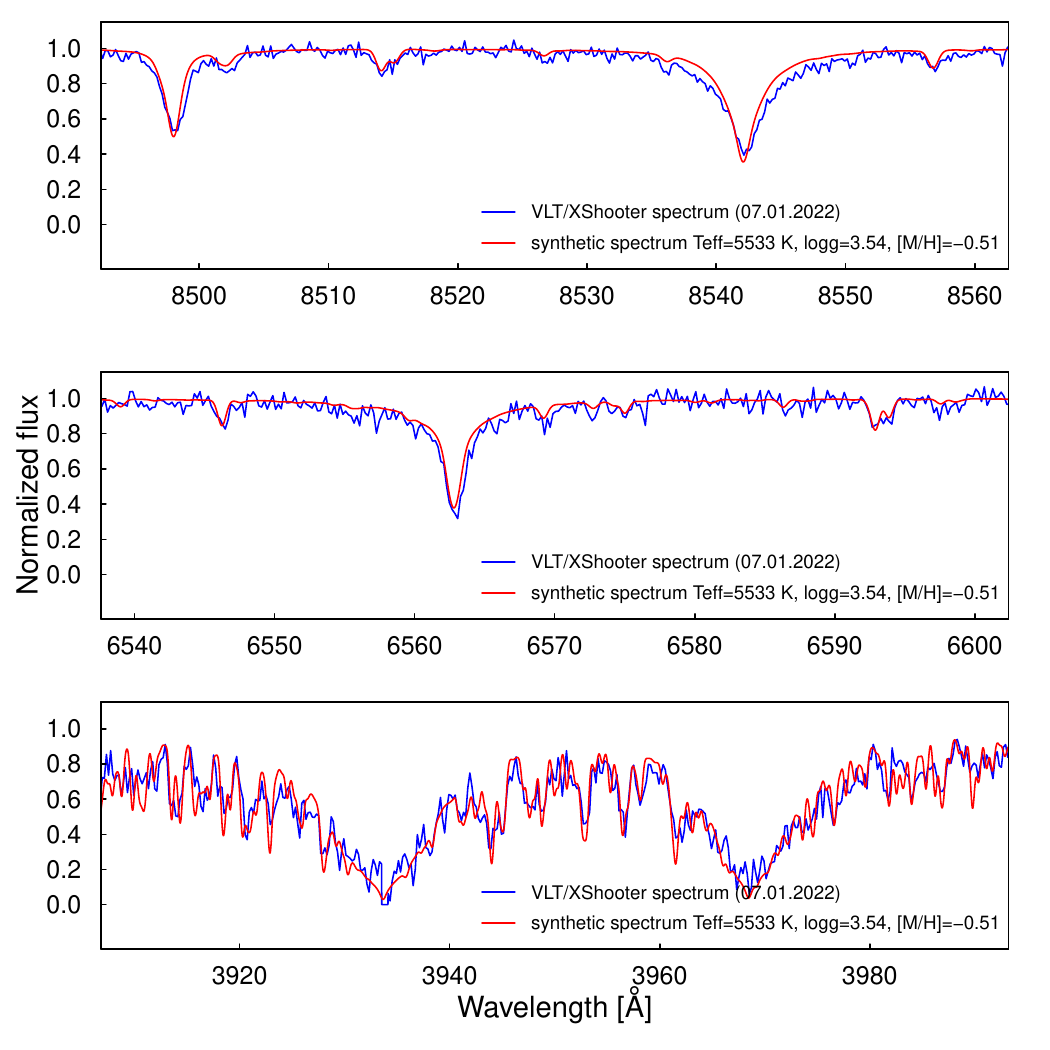}
    \caption{The VLT/X-Shooter spectrum (blue) and the best-matching fit of synthetic spectrum (red) generated for specific atmospheric parameters are presented. The Ca~II triplet ({\it top panel}), H${\alpha}$ line ({\it middle panel}) as well as Ca H and K lines ({\it bottom panel}) are visible.}
    \label{fig:xs-spectra}
\end{figure*}

The high-resolution spectrum from X-Shooter has been used for determining of the atmospheric parameters, i.e., effective temperature $T_{\rm eff}$, surface gravity $\log g$ and metallicity [M/H], of the source star. It was possible thanks to a spectral line fitting method by using {\it iSpec}\footnote{\url{https://www.blancocuaresma.com/s/iSpec}} package \citep{BlancoCuaresma2014, BlancoCuaresma2019}. We used the SPECTRUM\footnote{\url{http://www.appstate.edu/~grayro/spectrum/spectrum.html}} radiative transfer code to generate a set of synthetic spectra based on a grid of MARCS models \citep{Gustafsson2008}, solar abundances taken from \citet{Grevesse2007} and line list with atomic data from Gaia-ESO Survey (GESv6; \citealt{Heiter2021}). The GESv6 line list covers the wavelength range from 420 to 920~nm so this method uses only UVB and VIS part of the X-Shooter spectrum. The best-matching fit was found for the following parameters: $T_{\rm eff} = (5533 \pm 89)$~K, $\log g = (3.54 \pm 0.19)$, $\mathrm{[M/H]} = (-0.51 \pm 0.07)$~dex, and is presented together with the observational X-Shooter data in Fig.~\ref{fig:xs-spectra}.

Moreover, both spectra have been modeled with a template matching method using \texttt{Spyctres}, similarly to \citet{Bachelet2022,spyctres}. The latest version of Spyctres includes an update of the extinction law from \citet{Wang2019}. Briefly, we modeled the two spectra using the stellar template library from \citet{Kurucz1993} as well as the SED at the time of spectra acquisition, including the source magnification $A(\mathrm{t})$. This allows an accurate flux calibration and ultimately the estimation of $A_V$ and the stellar parameters. With this method, the final solution for the source star parameters was found as $T_{\rm eff} = (5297\pm30)$~K, $\log g = (3.50_{-0.25}^{+0.30})$, $\mathrm{[M/H]} = (-0.7_{-0.1}^{+0.3})$~dex, together with the line-of-sight extinction in V band: $A_V = 1.55_{-0.04}^{+0.03}$~mag.

\begin{figure*}
    \centering
    \includegraphics[width=0.95\textwidth]{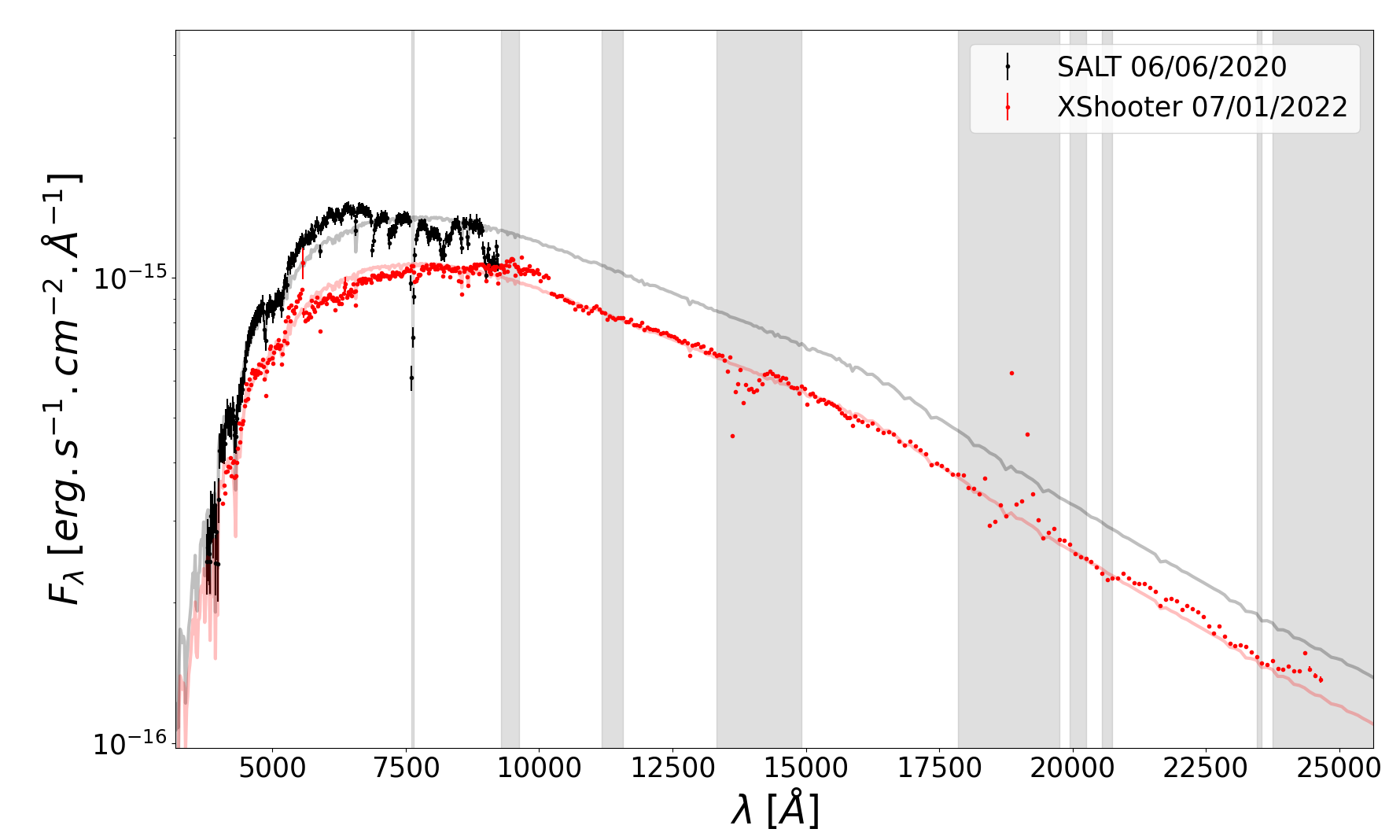}
    \caption{The two spectra from SALT (black) and X-Shooter (red), as well as the best models, are visible. The gray vertical lines indicates telluric bands, where the data were not used for the modelling.}
    \label{fig:spectra}
\end{figure*}

Both spectra and the results of template matching are visible in Figure~\ref{fig:spectra} while the modelling results are displayed in Table~\ref{tab:spectra}. The results obtained in spectroscopic analysis are in good agreement with the measurements from the GaiaDR3 release, where $T_\mathrm{eff}=(5434\pm16)~\mathrm{K}$, $\log g = (3.53\pm0.02) $, $[\mathrm{M}/\mathrm{H}]=(-0.520\pm0.001)~\mathrm{dex}$ (GaiaDR3 6054150372473485696; \citealt{Gaia2016,GaiaDR3}).

\begin{figure*}[h]
    \centering
    \includegraphics[width=0.9\textwidth]{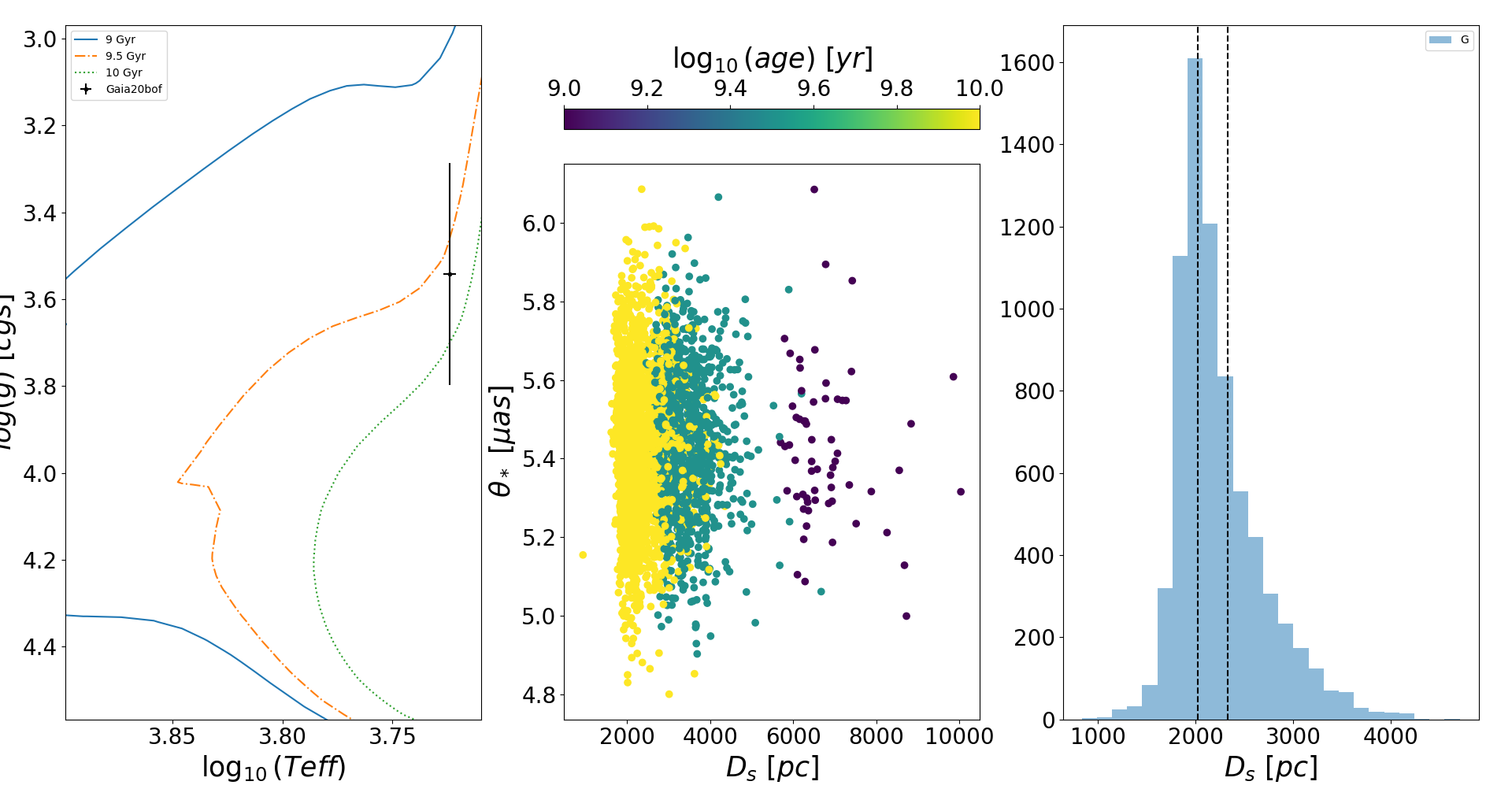}
    \caption{(Left)  PARSEC stellar isochrones for 9, 9.5 and 10 Gyr with a fixed metallicity of -0.5. The source is most likely an old subgiant. (Middle) Source angular radius based on the spectra models as a function of the source distance and age (color-coded). The points are derived from the template matching MCMC chains and the stellar isochrones. (Right) Source distance based on the Gaia source measurements from the model ($G_s\sim16$ mag) and the extinction estimated from the template matching modeling. Black dash lines indicates the 1-$\sigma$ confidence region from the parallax measurement of Gaia.}
    \label{fig:source}
\end{figure*}

\begin{table}

\setlength\extrarowheight{5pt}
    \centering

\caption{Summary of the source properties from the spectral analysis and GaiaDR3.}  
\label{tab:spectra}
\begin{tabular}{cccc}
\hline
\hline
Parameter & Template matching & Line fitting & GaiaDR3 \\
\hline
$A_V$ [mag] & $1.55_{-0.04}^{+0.03}$ & - & - \\
$T_\mathrm{eff}$ [K] & $5297\pm30$ & $5533\pm89$ & $5434\pm16$ \\ 
$\log g$ (cgs) & $3.50_{-0.25}^{+0.30}$ & $3.54\pm0.19$ & $3.53\pm0.02$ \\
$[\mathrm{M}/\mathrm{H}]$ [dex] & $-0.7_{-0.1}^{+0.3}$ & $-0.51\pm0.07$ &$-0.520\pm0.001$ \\
\\
\hline
\hline
\end{tabular}
\end{table}

\subsection{Distance to the source}
\label{sec:distance}
Using the derived parameters and the PARSEC stellar isochrones\footnote{\url{http://stev.oapd.inaf.it/cgi-bin/cmd}} \citep{Bressan2012}, we found that the source is most likely an old subgiant of G2 spectral type, as can be seen in Figure~\ref{fig:source}. Combining the source luminosity, the source apparent magnitude in the G band and the extinction law derived from spectral analysis, we found that the source angular radius $\theta_*= 5.4\pm0.2~\mathrm{\umu as}$ is independent of the source age and distance. We note that the angular radius value is also confirmed by the color-radius relation from \citet{Boyajian2014} with $\theta_*=5.2\pm0.2~\mathrm{\umu as}$ (see Table~\ref{tab:Models}).

Based on the stellar isochrones, the extinction obtained from the spectra fits and assuming the source magnitude $G=(16.00 \pm 0.05)~\mathrm{mag}$, we estimated the source distance to be $D_\mathrm{S}=2.1^{+0.5}_{-0.2}~\mathrm{kpc}$, as can be seen in the right panel of Fig.~\ref{fig:source}. We note that using different source magnitude bands (like SDSS-i') and models leads us to similar conclusion.

Moreover, assuming the typical absolute magnitude $M_{\rm V}= (3.0 \pm 0.3)$~mag \citep{Straizys1992} for metal-poor G2-type subgiant, we also calculated the distance to the source analytically. This way, we obtained the value of $D_\mathrm{S}=(1.95\pm0.35)~\mathrm{kpc}$.

These estimates are in excellent agreement with the parallax measurements from Gaia: $D_\mathrm{S}=2.33^{+0.03}_{-0.02}~\mathrm{kpc}$ (\url{distance_gspphot}). In addition, the value of the distance inferred for the Gaia20bof parallax published in GaiaDR2 \citep{Bailer-Jones2018} is $D_\mathrm{S}=4.32^{+2.94}_{-1.81}~\mathrm{kpc}$, while the updated value for the parallax published in GaiaEDR3 assuming geometric and photogeometric approach \citep{Bailer-Jones2021} is $D_\mathrm{S}=2.18\pm0.15~\mathrm{kpc}$ and $D_\mathrm{S}=2.14^{+0.15}_{-0.14}~\mathrm{kpc}$, respectively. All of these values, except the GaiaDR2 distance, are also in good agreement with the one obtained by us. Therefore, we decided to use the final source distance as $D_\mathrm{S}=2.1^{+0.5}_{-0.3}~\mathrm{kpc}$, from the spectra analysis. We note that at this distance, the galactic coordinates of the source are $(x,y,z)=(0.98^{+0.29}_{-0.15},-1.74^{+0.26}_{-0.52},-0.03^{+0.01}_{-0.00})$ kpc, that coincides with an overdensity of stars measured by Gaia \citep{Drimmel2023}.

\section{Future constraints}
\label{sec:future}
Because the source is close, $D_\mathrm{S}\sim 2~\mathrm{kpc}$, the lens is therefore closer. Moreover, the blend magnitudes also give an upper limit on the total mass of the lens $M_\mathrm{L} \le 0.8~M_\odot$ (and $D_\mathrm{L} \ge 0.4~\mathrm{kpc}$). 
Indeed, the combination of the source distance with the microlensing parallax $\pi_\mathrm{E} \sim 0.5$ from the models imply that any lens closer than $D_\mathrm{L} \le 0.2~\mathrm{kpc}$ would have been brighter than the observed blend. Therefore, the lens system is a low mass and relatively close binary, but its exact nature will be revealed with additional observations in the near future.

\subsection{Gaia astrometry}
\begin{figure}[h]
    \centering
    \includegraphics[width=1.25\textwidth,angle=90]{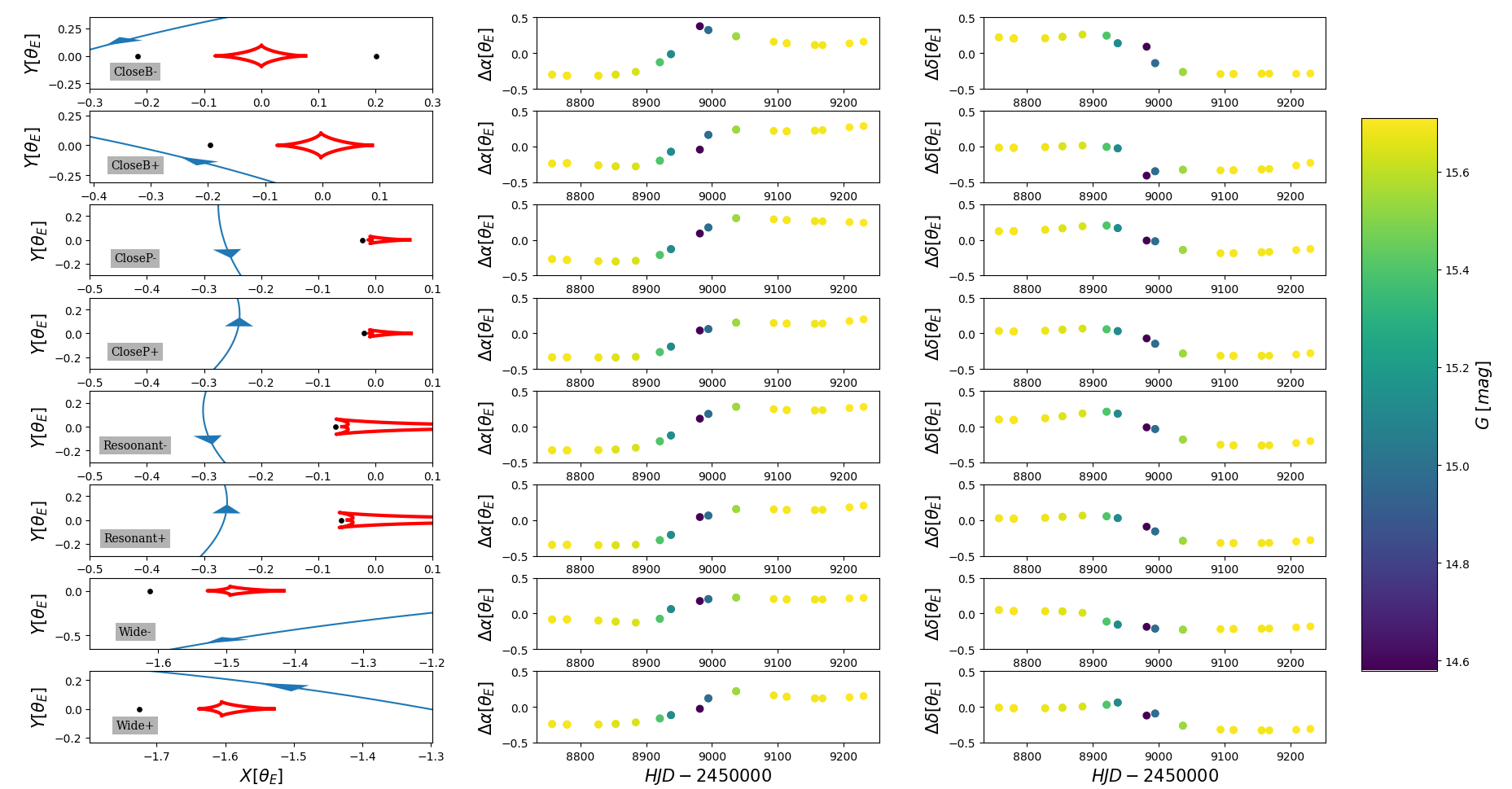}
    \caption{(Left) The source trajectories (blue), caustics (red) and critical curves (black) for the eight models. Note that the centre of mass of the lenses is kept fixed at (0,0). The microlensing astrometric deflections in right ascension (Middle) and in declination (Right) versus time are also displayed. The color indicates the observed Gaia G magnitude.}
    \label{fig:astrometry}
\end{figure}

The next data release from the Gaia Mission (DR4, $\sim$\,2025), will include astrometric time series that will help constrain the models. While the photometric lightcurves are almost identical for all of the models, the astrometric microlensing signals \citep{Walker1995,Dominik2000} can differ significantly. In  Figure~\ref{fig:astrometry} we show the astrometric microlensing signals as seen by Gaia for the eight models presented, assuming $\theta_\mathrm{E} = 1~\mathrm{mas}$. 
It is clear that the exquisite astrometric precision of Gaia will allow the selection of the most plausible model (or, at least, eliminate most of them) as well as measuring $\theta_\mathrm{E}$. 
This has already been done for the event Gaia16aye \citep{Wyrzykowski2020} \footnote{\url{https://www.cosmos.esa.int/web/gaia/iow\_20210924}}, which is similar in brightness to Gaia20bof, but presents more caustic crossing features. In the case of Gaia16aye, the early access to the astrometric time series confirmed both the microlensing model and the angular Einstein ring radius measurement of $\sim 3~\mathrm{mas}$.

\subsection{High resolution imaging}
By directly measuring the lens fluxes and/or the relative proper motions, high resolution imagers have been able to place strong constraints on microlensing systems, see for example \citet{Beaulieu2018,Bhattacharya2018,Vandourou2020}. In the case of Gaia20bof, high resolution imaging will also bring new constraints. First, the models predict different blend magnitudes in different bands, as listed in the Table~\ref{tab:Models}. A direct measurement of the lens flux, as well as a better analysis of the blend, will limit the number of possible scenarios. Moreover, the lens/source relative proper motions are significantly different between the models. Indeed, given that $\vec{\pi}_\mathrm{E} = \pi_\mathrm{E}\,\vec{\mu_{LS}}$ the (geocentric) relative proper motion can be written as \citep{Gould2004}:
\begin{equation}
{{\mathbf{\mu_{LS}}}\over{\theta_\mathrm{E}}} = {{1}\over{t_E}}{{\mathbf{\pi_{E}}}\over{\pi_E}}   
\end{equation}
All models predict different timescales $t_E$ and parallax vectors $\mathbf{\pi_E}$ (especially the Wide models). Therefore, 
observations of the lens and the source at different epochs will discriminate between models and ultimately provide a measurement of $\theta_E$ \citep{Vandourou2020}. Assuming a typical value of $\theta_E\sim1$ mas, high-resolution observations could start in about 5 years, with a lens and source separation $\delta\ge10$ mas for all models except Wide-.

\subsection{Radial velocity measurements}
As soon as the lens and source separate, it will be possible to conduct radial-velocity follow-up of the host that should lead to the full characterization of the system in combination with the two methods described previously. The radial-velocity monitoring of microlensing lenses is challenging, due to the faintness of the host. However, it has been done in the past on at least two occasions. \citet{Yee2016} used the Keck High-resolution Echelle Spectrometer \citep{Vogt1994} and the Magellan Inamori Kyocera Echelle spectrometer \citep{Bernstein2003} to measure the radial velocity signal of the host of the microlensing event OGLE-2009-BLG-020 \citep{Skowron2011}. Their data confirms and refine the stellar binary parameters from the original publication. A second radial velocity test has been made by \citet{Boisse2015} on the microlensing predictions of OGLE-2011-BLG-0417 \citep{Shin2012}. In this case, \citet{Boisse2015} did not measure any modulations in the radial velocity of the host, indicating strong tensions with the microlensing models later confirmed by high-resolution imaging \citep{Santerne2016}. Ultimately, \citet{Bachelet2018} identified a competitive microlensing model that explains the lack of modulations. In the case of Gaia20bof, radial velocity observations will be challenging, as the lens cannot be brighter than the measured blend ($ib\le18$ mag). However, it could be done with the most recent spectrographs, such as ESO ESPRESSO \citep{Pepe2021}.

\section{Conclusions}
\label{sec:conclusion}
We report the detection of a close ($\le 2~\mathrm{kpc}$) binary lens by combining space-based and ground-based time series photometry. Spectroscopic data indicates that the source is a relatively close ($D_\mathrm{S}\le 2~\mathrm{kpc}$) red sub-giant, in good agreement with parallax measurements from Gaia. The photometric lightcurves can be explained by a binary lens model. However, several degenerate models can reproduce the observations. We discuss several methods to distinguish between the models in the future. First, high-resolution imaging can confirm or reject the model prediction on the relative proper-motion direction, as well as give a measurement of $\theta_E$. Indeed, with a lens distance $D_\mathrm{L}\le 2~\mathrm{kpc}$, it is almost certain that the lens will be observable with current facilities such as Keck. High-resolution images would also provide a measurement of the host flux that will need to be compared with the blend values reported for each model. Secondly, the Gaia astrometric time series, expected with the Gaia DR 4, would be extremely useful for constraining the lens properties. Indeed, the eight models predict significantly different astrometric shifts that should be measurable if $\theta_\mathrm{E}$ is not too small (i.e. $\theta_\mathrm{E} \ge 0.5~\mathrm{mas}$). In this case, the astrometric time series will allow the estimation of the lens mass via the direct measurement of $\theta_E$. Finally, we discussed the possibility of measuring the radial velocity of the host. This would be challenging, as we expect the host to be faint, but could be done with the most precise instruments.

\begin{acknowledgements}
EB gratefully acknowledge support from NASA grant 80NSSC19K0291. EB's work was carried out within the framework of the ANR project COLD-WORLDS supported by the French National Agency for Research with the reference ANR-18-CE31-0002. This work was authored by employees of Caltech/IPAC under Contract No. 80GSFC21R0032 with the National Aeronautics and Space Administration. This research has made use of the NASA Exoplanet Archive, which is operated by the California Institute of Technology, under contract with the National Aeronautics and Space Administration under the Exoplanet Exploration Program.  This work has made use of data from the European Space Agency (ESA) mission
{\it Gaia} (\url{https://www.cosmos.esa.int/gaia}), processed by the {\it Gaia}
Data Processing and Analysis Consortium (DPAC,
\url{https://www.cosmos.esa.int/web/gaia/dpac/consortium}). Funding for the DPAC
has been provided by national institutions, in particular the institutions
participating in the {\it Gaia} Multilateral Agreement. This research has made use of the VizieR catalogue access tool, CDS, Strasbourg, France. 
YT acknowledges the support of DFG priority program SPP 1992 “Exploring the Diversity of Extrasolar Planets” (TS 356/3-1).
This work is supported by Polish NCN grants: Daina No. 2017/27/L/ST9/03221, Harmonia No. 2018/30/M/ST9/00311 and MNiSW grant DIR/WK/2018/12. 
The BHTOM project has received funding from the European Union's Horizon 2020 research and innovation programme under grant agreement No. 101004719 (OPTICON RadioNet Pilot, ORP).
BHTOM acknowledges the following people who helped with its development: 
Patrik Sivak, Kacper Raciborski, Piotr Trzcionkowski and AKOND company.
This paper made use of the Whole Sky Database (wsdb) created by Sergey Koposov and maintained at the Institute of Astronomy, Cambridge by Sergey Koposov, Vasily Belokurov and Wyn Evans with financial support from the Science \& Technology Facilities Council (STFC) and the European Research Council (ERC), with the use of the Q3C software 
(
http://adsabs.harvard.edu/abs/2006ASPC..351..735K
). RFJ acknowledges support for this project provided by ANID's Millennium Science Initiative through grant ICN12\textunderscore 009, awarded to the Millennium Institute of Astrophysics (MAS), and by ANID's Basal project FB210003.

DAHB acknowledges support from the South African National Research Foundation. PAW acknowledges financial support from the NRF (grant no. 129359) and UCT.
The SALT observations were obtained under the SALT Large Science Programme on transients (2018-2-LSP-001; PI: DAHB). Polish participation in SALT is funded by grant number MEiN nr 2021/WK/01.

\end{acknowledgements}

\bibliographystyle{aasjournal}  
\bibliography{Gaia20bof.bib}

\end{document}